\documentclass[conference]{IEEEtran}
\IEEEoverridecommandlockouts
\usepackage{cite}
\usepackage{amsmath,amssymb,amsfonts}
\usepackage{algorithmic}
\usepackage{graphicx}
\graphicspath{{media/}}
\usepackage{textcomp}
\usepackage{xcolor}
\def\BibTeX{{\rm B\kern-.05em{\sc i\kern-.025em b}\kern-.08em
    T\kern-.1667em\lower.7ex\hbox{E}\kern-.125emX}}
\begin{document}

\title{Crest factor optimization of multi-tone signals generated for
low-temperature sensor readout\\
}

\author{\IEEEauthorblockN{L.~H.~Arnaldi}
\IEEEauthorblockA{\textit{Laboratorio Detección de Partículas y Radiación} \\
\textit{Bariloche Atomic Center and Instituto Balseiro, National
Commission of Atomic Energy}\\
S.~C.~de~Bariloche, RN, Argentina \\
arnaldi@cab.cnea.gov.ar}
}

\IEEEpubid{\begin{minipage}{\textwidth}\ \\[12pt]\centering\scriptsize
\copyright\ 2023 IEEE. Personal use of this material is permitted. Permission from IEEE must be obtained for all other uses, in any
current or future media, including reprinting/republishing this material for advertising or promotional purposes, creating new
collective works, for resale or redistribution to servers or lists, or reuse of any copyrighted component of this work in other
works. Published in: 2023 Argentine Conference on Electronics (CAE), DOI: 10.1109/CAE56623.2023.10086985.
\end{minipage}}
\maketitle


\date{\today}

\begin{abstract}
	Microwave kinetic inductance detectors (MKID) are a growing technology of
	cryogenic detectors and have a great potential in radio-astronomy and cosmic
	microwave background (CMB) characterization. A review of the general
	readout performance requirements for experiments requiring the multiplexing of
	large arrays ($\sim$10.000 pixels) of such cryogenic detectors is
	presented. The focus is on the crest factor optimization process for proper
	excitation and readout of these large arrays of detectors. It is shown an
	algorithm to optimize the phase generation of the carriers to obtain a low
	crest factor, favoring the dynamic range usage of the electronic chain.
\end{abstract}


\section{Introduction}\label{intro}
Developing the next generation of cryogenic detectors for CMB and sub-millimeter
wavelength studies requires densely instrumented arrays of sensors to meet its
scientific goals. Reading large quantities ($\geq$10.000 per camera) of
sub-Kelvin sensors, for example, as those proposed for the
QUBIC~\cite{Battistelli2011} and
CMB-S4~\cite{Abitbol2017} experiments, will require substantial improvements in
cold and warm readout techniques. To reduce the readout cost per sensor and the
complexity of integration, efforts are currently focused on achieving higher
multiplexing density, while keeping the readout noise below the intrinsic
detector noise and presenting manageable thermal loads. Highly-multiplexed cold
readout technologies in active development include microwave kinetic inductance
(MKID)~\cite{Day2003, Mazin2004, Gao2008t}, transition edge sensors
(TES)~\cite{Piat2021} and microwave rf-SQUID~\cite{Irwin2004,Kempf2017} sensors.
Two of them, the MKID and the rf-SQUID, exploit the high quality factors of
superconducting microwave resonators to densely channelize sub-Kelvin sensors
into the bandwidth of a microwave transmission line. 

The main advantage of the working principle of these cryogenic detectors is that
it is relatively easy to perform the frequency division multiplexing (FDM)
required for its excitation and readout, and this enables the implementation of
large arrays of this type of sensors.  In these systems, however, it is
necessary to use the largest possible carrier amplitude without exceeding the
dynamic range of the digital to analog converter (DAC), to maximize the
signal-to-noise ratio (SNR) of the exciting waveform. This implies that the
crest factor must be as low as possible.


This paper reviews the general readout performance requirements for experiments
requiring the multiplexing on the order of 10.000 or more MKIDs (pixels) and
addresses the necessary crest factor optimization process for proper excitation
and readout of these large arrays of detectors. The focus will be put on MKIDs
detectors as they are still a growing technology and have a great potential in
radio-astronomy and CMB
characterization~\cite{Walter2020,Araujo2014a,Zobrist2021,Ulbricht2021}. 

\IEEEpubidadjcol   
\section{Readout electronics specification}\label{especificaciones_de_disenio}
\begin{figure*}[!ht]
	\centering
	\includegraphics[width=\textwidth]{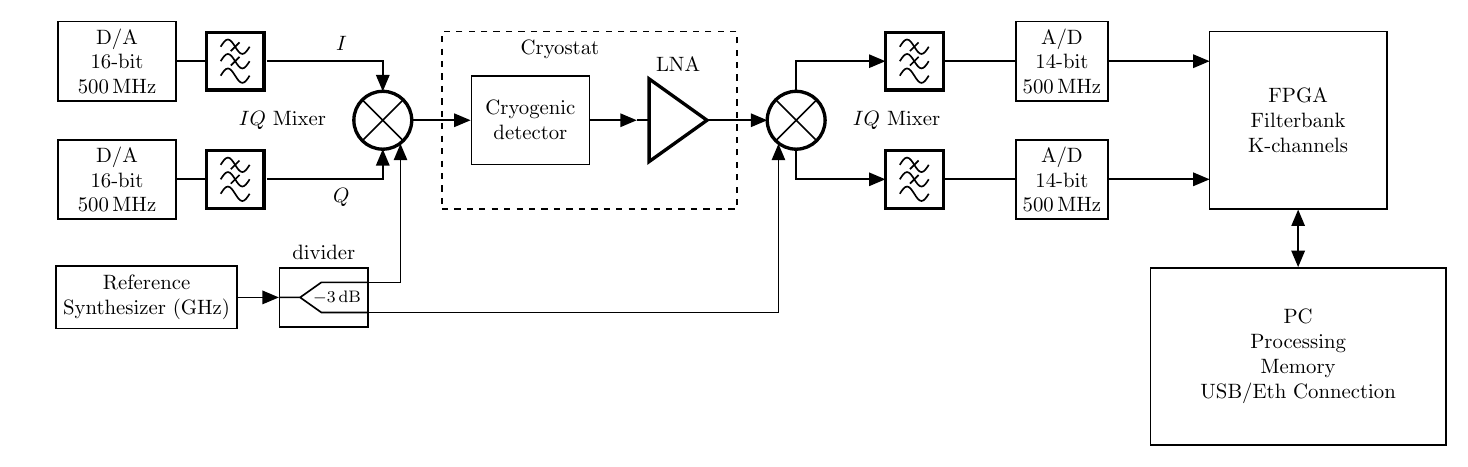}
	\caption{
		Block diagram of a proposed system for the readout of
		superconducting detectors working at microwave frequencies. For clarity,
		various amplifiers and attenuators that make up the drive and read chains of
		the MKIDs have been left out. 
		}
	\label{fig_readout_general}
\end{figure*}

\subsection{General considerations}\label{general_issues}
Reading an array of MKIDs is done with electronics at room temperature, a huge
advantage compared to other sensing technologies, such as transition edge
sensors, which require cryogenic SQUIDs for each pixel. By using fast A/D
converters to digitize high-bandwidth signals, the flexibility of FPGAs, coupled
with advanced signal processing techniques, such as the algorithms discussed in
this paper, can be used to extract the desired signals.

The readout electronics have the general task of performing multiple complex
microwave transmission ($S_{21}$) measurements in real time, around 4-10\,GHz,
to monitor the instantaneous resonance frequency and dissipation of the
superconducting micro-resonators working as detectors in the specific range of
wavelengths of the experiment. One of the most stringent requirements for the
electronic readout system is that it should be able to perform the measurements
of the parameter $S_{21}$ at a noise level that is limited only by the noise
temperature of the cryogenic amplifier (LNA) that follow the arrangement of
MKIDs. This amp will generally be a high electron mobility transistor (HEMT)
type.  Therefore, the noise contributed by electronic components at room
temperature should be below this level.  Typically, the noise/carrier ratio at
the output of the HEMT amplifier is expected to be on the order of
-100\,dBc/Hz~\cite{McHugh2012}. 
The noise contribution from the readout electronics must be below the HEMT
noise. In other words, the noise/carrier ratio at the output of the HEMT
amplifier must be preserved through the electronic chain of the readout system.



\figurename~\ref{fig_readout_general} shows a block diagram for a MKID readout
system. A read circuit for a single MKID resonator would be configured as
follows: a frequency synthesizer provides a signal at the resonator frequency
(4-10 GHz). This reference signal is then divided by means of a matched power
divider. One copy, the drive signal, is sent through the cryostat, resonator,
and amplifiers, past the radio frequency (RF) port of the up-conversion IQ
mixer. The other copy is used as input to the local oscillator (LO) of the
downconverting IQ mixer. 
The output of the down-conversion mixer provides the in-phase (I) and quadrature
(Q) portions of the signal, from which the amplitude and phase changes
introduced by the resonator can be derived.

Since the off-resonance transmission of the MKIDs is $\simeq 1$, the reading
of a large number of resonators can be multiplexed using a single drive signal
containing a comb of frequency tones tailored to each MKID. This comb will be
generated by fast dual D/A converters that will drive the $IQ$ ports of an
upconversion mixer. The multiplexed drive signal will be generated at the
baseband frequency 
using high-speed,
high-bandwidth D/A converters. These D/A converters are used to reproduce a
predefined and easily modifiable waveform, which will be stored in memory. The
requirement for easy waveform modification arises from the fact that the
resonant frequencies of the MKIDs change as a function of the observation
conditions~\cite{Noroozian2009a, Czakon2009a}. The output signals from the
baseband D/A converter are converted to the microwave band and sent to the
cryostat to drive the resonators. By using a dual DAC scheme with an up
converting $IQ$ mixer, and a similar scheme with dual ADCs on the output side,
the need for a single sideband conversion is avoided. A firmware will then be
used in the FPGA to digitally process this data, essentially demodulating it at
each frequency of the comb to determine the response of each resonator. The
demodulated signal for each resonator can then be stored in external memory at a
data rate 
depending on the application.

\subsection{The FPGA firmware}
In general, an FPGA is necessary to demodulate and decimate the output signal of
the array of MKIDs, without adding noise exceeding the noise of the ADC. 
In Section~\ref{c3_generacion_tonos} the whole process of generation in the FPGA of
the spectral components for the excitation of the MKIDs is analyzed in greater
detail.

The LO inputs, for both the up and down mixers, are controlled with the same
microwave synthesizer at the frequency $f_\text{sm}$. The D/A converters
used to generate the input signal are programmed to reproduce a stored frequency
comb as follows:
\begin{subequations}\label{eq_tonos_DAC}
	\begin{align}
		S_I(t) &= \sum_k A_k \cos(2\pi f_k t + \phi_k)\\
		S_Q(t) &= \sum_k A_k \sin(2\pi f_k t + \phi_k)
	\end{align}
\end{subequations}
Here $k$ is the index for the read tones, $k = 1, \ldots, N$, for $N$
total resonators in the array, and $t$ is the time. $A_k$, $f_k$ and $\phi_k$
represents the amplitude, frequency and phase, respectively, of the
$k$-th reading tone. $\phi_k$ is relative to $t = 0$ (the beginning of the
memory that provides the tones). In general, the phases $\phi_k$ should be
chosen in a way that mantain the output power of the D/A converters
approximately constant, favoring a low crest factor.
$A_k$ is the amplitude at the output of the D/A converter. This will be on the
order of $2^{B-1}/\sqrt{N}$, where $B$ is the number of bits in the D/A converter. 
The frequency $f_k$ must be in the Nyquist range,
		\begin{equation}\label{eq:f_DAC}
			-f_\text{DAC}/2 < f_k < f_\text{DAC}/2
		\end{equation}
		where $f_\text{DAC}$ is the sample rate of the DAC. This frequency is also
		an integer multiple of $1/T$, where $T$ is the length (in
		time) of the tone memory for the DAC. The tone memory contains $T\times
		f_\text{DAC}$ samples for each of the $I$ and $Q$ inputs.
The upconversion mixer gets $S_I$ and $S_Q$ as inputs, along with the
microwave LO signal at $f_\text{sm}$, and outputs the input signal to the
cryostat:
\begin{equation}\label{eq:f_DAC}
	V_{in} (t) = S_I (t) \cos(2\pi f_\text{sm} t) - S_Q
	(t) \sin(2\pi f_\text{sm} t),
\end{equation}
plus some global scale factor due to converter efficiency which is not taken
into account here for convenience and simplicity of reasoning. It can be shown
that the above equation is equivalent to
\begin{equation}\label{eq:f_DAC2}
	V_{in} (t) = \sum_k A_k \cos(2\pi(f_\text{sm} + f_k) t + \phi_k ),
\end{equation}
so that the generated frequencies are in the range $[f_\text{sm}-f_\text{DAC}/2,
f_\text{sm} + f_\text{DAC}/2]$. After being sent through the
detector array, the phases and amplitudes of the signals are modified, and the
signal coming out of the cryostat is:
\begin{equation}\label{eq_f_DAC3}
	V_{out} (t) = \sum_k R_k (t) \cos(2\pi(f_\text{sm} + f_k) t + \theta_k (t)).
\end{equation}
The goal here is to measure the quantities $R_k(t)$ and $\theta_k(t)$ for
all resonators. These two magnitudes represent the change in the amplitude and
phase of the signals, due to the change in the resonance frequency of the MKID
detector, resulting from the absorption of radiation.

The equation~\ref{eq_f_DAC3} can be more conveniently written in the form
\begin{equation}\label{eq_f_DAC4}
	c_k(t) \equiv R_k (t) e^{j\theta_k (t)},
\end{equation}
that is, $ c_k $ is the complex amplitude of the $k$-th tone. $c_k(t)$ is
expected to vary slowly with time\footnote{In this context, ``slow variation''
means on timescales such that $t \gg 1/f_k$.}, and for typical MKID
applications, we want to measure the variation on the time scales $\tau \sim
1/(100\,\text{Hz})\simeq 10\,\text{ms}$~\cite{Baselmans2012,Mazin2009}.

On the receiving side, another mixer $IQ$ is used to convert $V_{out}$ to
baseband. After anti-alias filtering and digitization, the result is
\begin{subequations}\label{eq_f_DAC5}
	\begin{align}
		I(t) &= \sum_k \frac{R_k (t)}{2} \cos(2\pi f_k t	+ \theta_k (t))\label{I_t}\\
		Q(t) &= -\sum_k \frac{R_k (t)}{2} \sin(2\pi f_k t + \theta_k (t))\label{Q_t}.
	\end{align}
\end{subequations}
These signals can be treated as complex signals
\begin{equation}\label{eq:f_DAC6}
	C(t) \equiv I(t)-jQ(t) = \frac{1}{2}\sum_k c_k (t) e^{j2\pi f_kt},
\end{equation}
where $c_k$ is the complex amplitude of the tone $k$, as defined in the
equation~\eqref{eq_f_DAC4}. Then an algorithm is used in the FPGA to extract
$\langle c_k \rangle $ from the complex data stream, at a sample rate of $1/T_
{out} \sim 100\,\text{Hz}$, slow enough so that the data can be
transmitted continuously and stored on a memory disk.

The amplitude for the decimated output sample can be written as
\begin{equation}\label{eq:f_DAC7}
	\langle c_k(s) \rangle =  \sum_t C(t)h(sT_{out}-t) e^{-j2\pi f_kt}
\end{equation}
where, in general, $-\infty<t<sT_{out}$, and the coefficients $h$
define the decimation filter. The fundamental task of the FPGA will then be to
perform the operation of the equation~\eqref{eq:f_DAC7} on the complex
signal of the equation~\eqref{eq:f_DAC6}.

\section{The tone generation process}\label{c3_generacion_tonos}
A large number of low-temperature detectors can be read through a few wires
using frequency division multiplexing. In this scheme, different carrier
frequencies are assigned to the MKIDs, and each MKID is part of a resonant
circuit $ LC $ tuned to pass only the power at its assigned frequency. Each
carrier is configured to be in the center of the resonator passband. Changes in
the surface impedance of the MKID modulate the amplitude and phase of its
carrier.

Because each MKID has its own bandwidth, the feeds of numerous MKIDs can be
carried on a single pair of cables. Similarly, all modulated signals from
numerous MKIDs can be transported via a single pair of cables to a HEMT
cryogenic amplifier. However, combined signals sometimes produce large peak
amplitudes on feed or read. This overloads the capabilities of the electronics.
Thus, it is necessary to develop an efficient algorithm to optimize the phases
and frequencies of the carriers to obtain a low crest factor ($CF$). 

In order to approximate our digital signals to \eqref{eq_f_DAC5} we can start
by analizing a real-valued, constant-amplitude signal. This approximation will
serve to understand the requirements for the algorithm to develop. If we
consider the digital multi-tone signal
\begin{equation}\label{eq_senial_multitono}
	u(n) = \sqrt{\frac{2}{N}}\sum_{k=N_0+1}^{N_0+N}\cos(kn+\theta_k)
\end{equation}
where $n$ is integer, $N_0 \geq 0$ and $N$ is the number of tones present
in the signal. The Fourier coefficients of the $u$ signal satisfies
\begin{equation}\label{eq_senial_multitono_fourier}
	|\hat{u_k}|^2 = \left\{\begin{array}{rcl}
		\frac{2}{N}, && N_0 < k \leq N_0 + N\\
		0, &&\text{otherwise}.
	\end{array}\right.
\end{equation}
Therefore, the effective value (see the Appendix for details) of $u$ is $||u||_2
=\sqrt{\frac{1}{2}\sum|\hat{u}_k|^2} = 1$, regardless of the phases $\theta_k$. For
this reason, $u$ is called a \emph{unity power signal} with a perfect band-pass
spectrum (if $N_0 > 0$) or a perfect low-pass spectrum (if $N_0 =
0$)~\cite{Boyd1986}. This is valid even for the complex signals in
\eqref{eq_f_DAC5}.

\subsection{Phase optimization in order to minimize the crest factor}\label{opt_fases}
While the rms value is independent of the phases, the peak (which is also
$CF(u)$) changes dramatically with the phases. The question to answer then is:
how should the phases $\theta_k$ be chosen in \eqref{eq_senial_multitono} to
minimize the crest factor of $u$, or at least to make it small?.

The worst choice of phases is $\theta_k = 0,\, k = 1, \ldots, N$, which produces
a peak of $||u||_\infty =\sqrt{2N}$ (which occurs at $n = 0$).

\begin{figure}[!ht]
	\centering
\hspace*{-0,50cm} \includegraphics[width=0.56\textwidth]{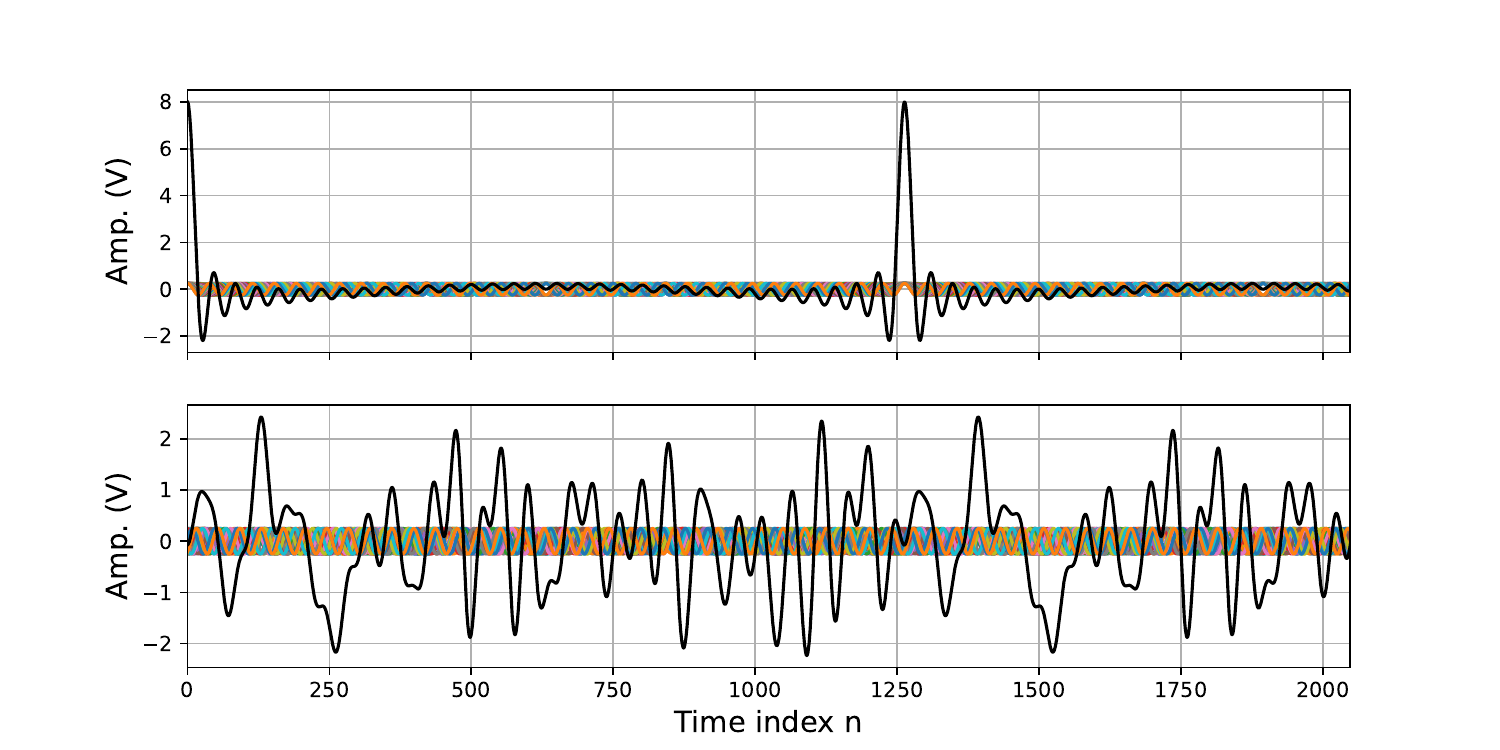}
	\caption{
		Signals, $u(n)$, with different crest factors. Both graphs show the FDM signal
		obtained as the sum of 32 tones (black). Tones have constant amplitudes and
		phases $\theta_k = 0$ (top) and constant amplitudes and random phases
		(bottom). The crest factor is 18.090\,dB for the constant case and 7.713\,dB
		for the random case.
		}
	\label{fig_fase_aleatorio_no_aleatorio}
\end{figure}

\figurename~\ref{fig_fase_aleatorio_no_aleatorio} shows a 32-tone multi-tone signal
without and with the addition of random phases. If the phases $\theta_k$ vary
linearly, for example $\theta_k = \tau n$, then the peak remains at $\sqrt{2N}$.
It is clear that a phase pattern chosen with some criteria is needed so that
cancellations always keep $|u(n)|$ small. In fact, in~\cite{Gersho1979} has been
shown that if the phases are chosen randomly, spikes on the order of
$\sqrt{\log N} $ occur, which is much smaller than $\sqrt{2N}$, but still grows
with $N$.

In the literature there are few articles that address the problem, for
example~\cite{Lindeman2012} investigates how to obtain a low crest factor for
the excitation of TES detector arrays at low carrier frequencies. The authors
mention that the lowest crest factor is obtained by applying carriers equally
spaced in frequency, which would not be the case for MKIDs, since as mentioned
above, the final resonance position depends on several factors in a MKID.
In~\cite{Boyd1986} two cases of phase generation for carriers are analyzed, one
due to Rudin~\cite{Rudin1959} and Shapiro~\cite{Shapiro1952} and another method
attributed to Newman~\cite{Newman1965}. Both methods are relatively simple to
implement and produce a nearly minimal crest factor for a multi-tone signal.
However, the author remarks that standard optimization techniques do not help
much in the design of low crest factor signals. The crest factor is a very
complicated function of the phases, with many local minima.
Finally, the study shows that the Newman phases produce a crest factor only one
and a half decibel lower ($\sim 4.6\,\text{dB}$) than those of Rudin and
Shapiro ($\sim 6\,\text{dB}$).

In this paper, Newman's tone generation method is analyzed, since it is the
method that allows obtaining the lowest $CF$ and by means of which crest
factors of $\sim 4.6\,\text{dB}$ are obtained for $N$ on the
order of a few hundred, decreasing even for larger $N$.

\subsubsection{The Newman's phases}\label{fases_newman}
To analyze the tone generation with Newman's method, the following phases are
defined for $N$ tones
\begin{equation}\label{eq_senial_multitono_NEW}
	\theta_k^{(New)} = \frac{\pi (k-1)^2}{N}.
\end{equation}
Then, we can define the multi-tone signal,
\begin{equation}\label{eq_senial_multitono_NEW}
	u_N^{(New)}(n) \triangleq
	\sqrt{\frac{2}{N}}\sum_{k=1}^{N}\cos((k+N_0)n+\theta_k^{(New)}).
\end{equation}
\begin{figure}[!ht]
	\centering
\hspace*{-0,5cm}	\includegraphics[width=0.56\textwidth]{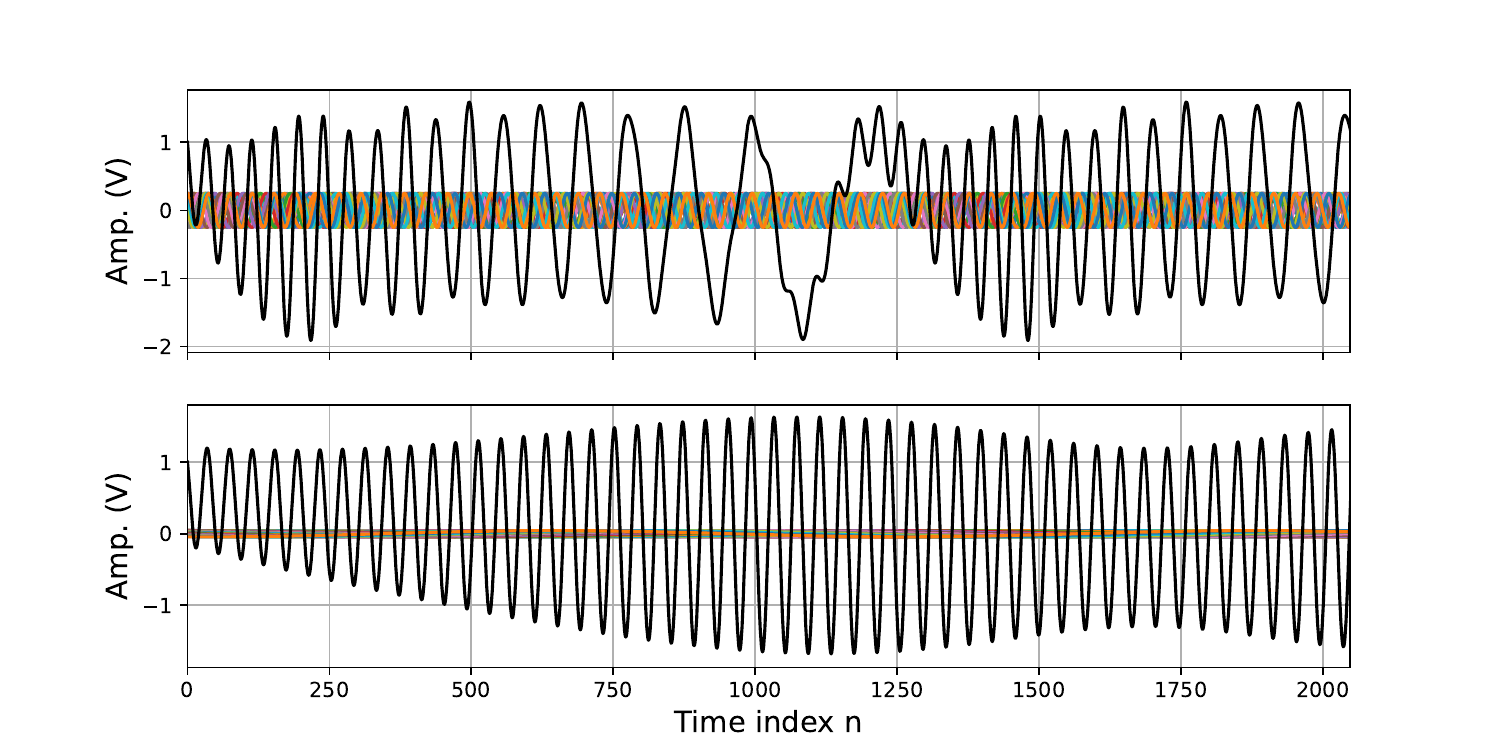}
	\caption{
		Multi-tone signals, $u_N^{(New)}(n)$, obtained by using different values of
		$N$, $N=32$ (top) and $N=1024$ (bottom), and constant amplitudes. 
	 }
	\label{fig_fases_tres_valores_N}
\end{figure}

\figurename~\ref{fig_fases_tres_valores_N} shows the multi-tone signals,
$u_N^{(New)}(n)$, obtained for two values of $N$. We have $CF=5.635$\,dB and
$CF=4.628$\,dB for $N=32$ and $N=1024$ tones, respectively.

\begin{figure}[!ht]
	\centering
\hspace*{-0,5cm} \includegraphics[width=0.56\textwidth]{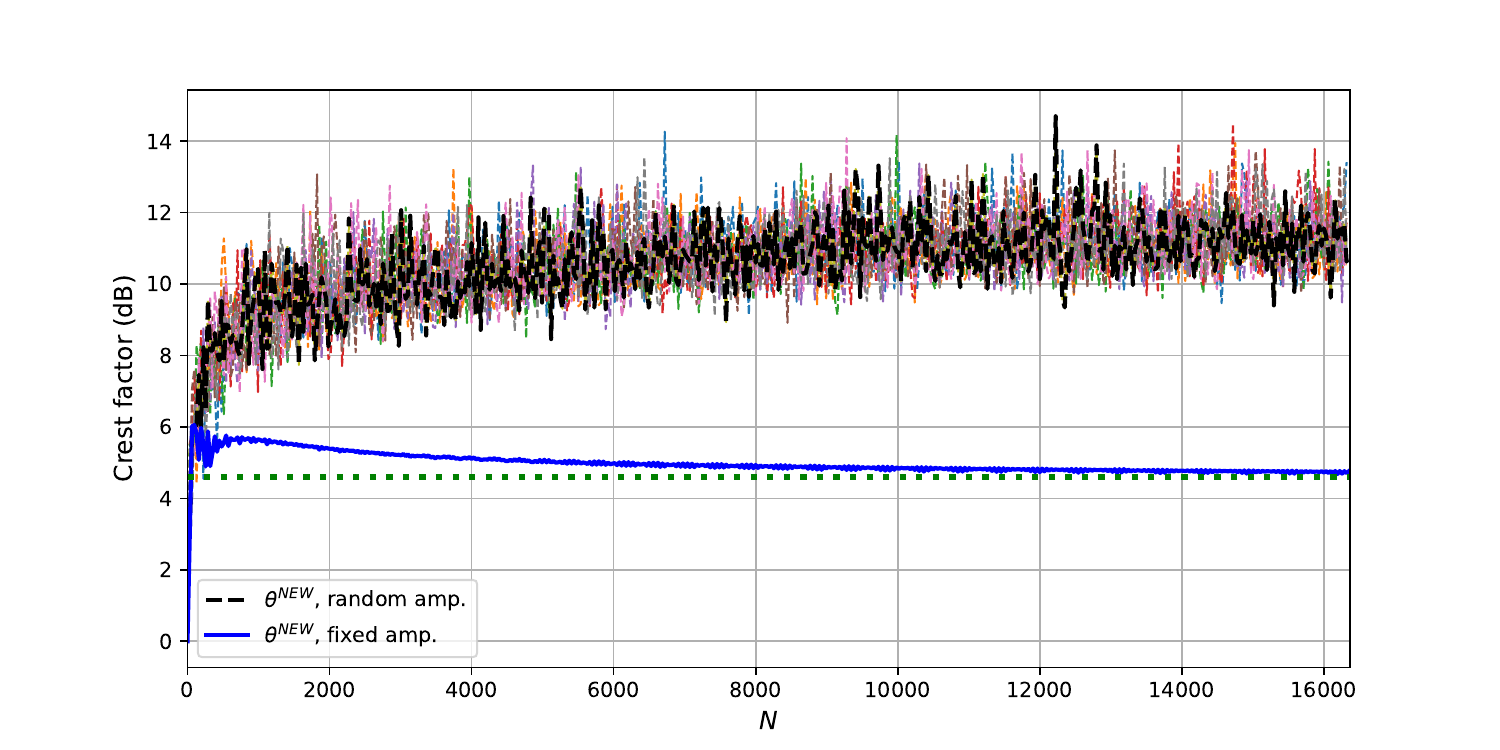}
	\caption{
		Evolution of the crest factor with Newman's phases for each $N$, up to $N =
		16384$. Through several runs, it is observed that there are no benefits to
		making the amplitudes random. The crest factor worsens for this
		case and there is a difference of $\sim 6\,\text{dB}$ with the case
		of Newman's generation of phases and constant amplitudes. The dotted line
		indicates the value of $CF=4.6\,\text{dB} $, the theoretical minimum
		achievable with the Newman phases.
		}
	\label{fig_fc_newman}
\end{figure}

In addition, it is worth wondering what happens if the amplitudes of the signals
defined by the equation~\eqref{eq_senial_multitono_NEW} are also changed, since
it is common practice in MKID readout systems to adjust the power level applied
to each resonator to optimize the response. To check the behavior of the crest
factor in this case, simulations were carried out applying random amplitudes and
in all cases it was found that the crest factor grows with $N$, tending to
stabilize at $\sim 10.5\,\text{dB}$. From these results, it can be stated that
in the case of random variation of phases and amplitudes, the peak of the FDM
signal is in the order
\begin{equation}\label{eq_fc_aleatorios}
	||u||_\infty = \left\{\begin{array}{rcl}
		2.6 \sqrt{\log{N}}, && N \geq 1000\\
		2.4 \sqrt{\log{N}}, && N < 1000.
	\end{array}\right.
\end{equation}
\figurename~\ref{fig_fc_newman} shows $CF(u_N^{(New)})$ for $N=1,\ldots, 16384$.
Numerical analysis shows that the crest factor for Newman's multi-tone signal is
always very small, about $4.6\,\text{dB}$ for a moderate $N$ (up to a few
hundred) and decreases very little for larger $N$.

\begin{figure*}[!ht]
	\centering
	\includegraphics[width=0.8\textwidth]{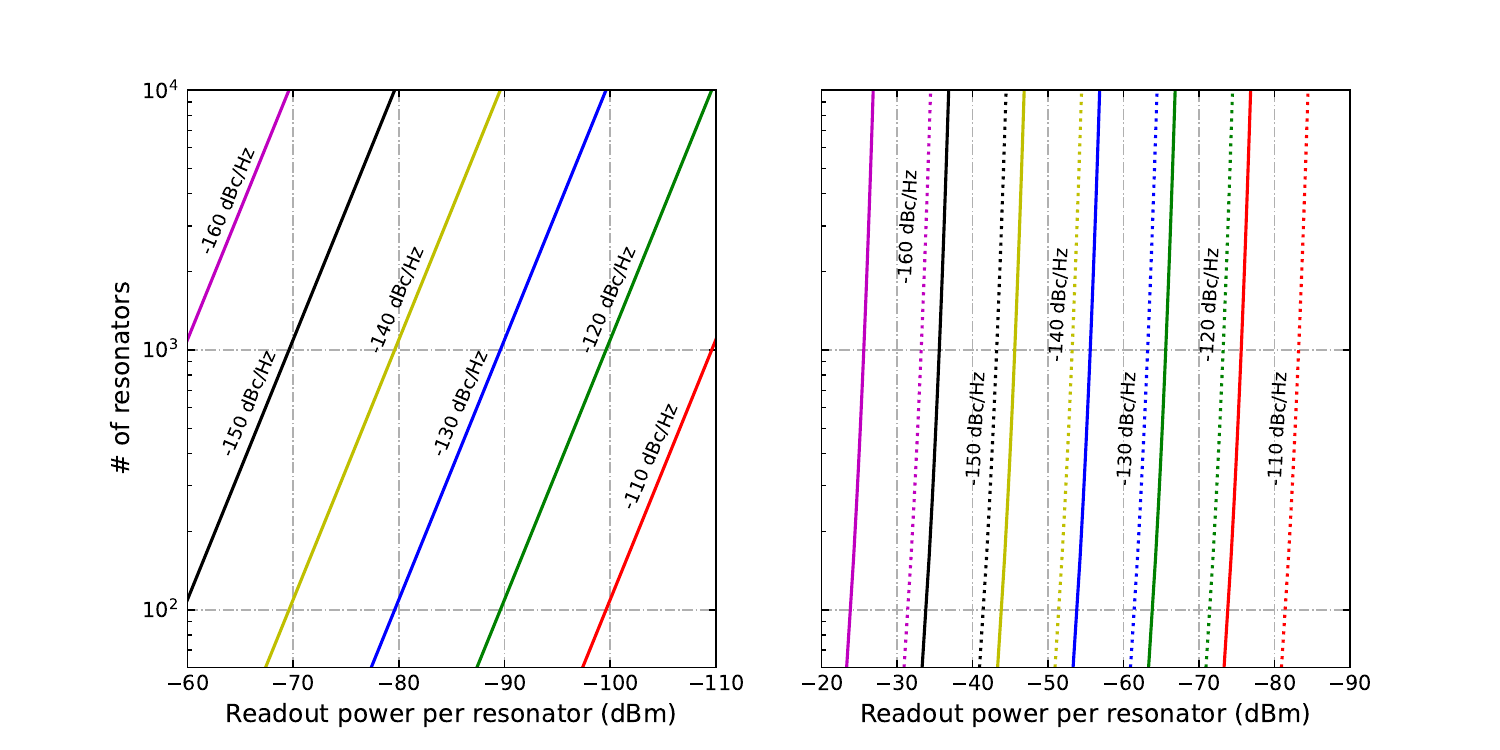}
	\caption{
		Required ADC noise density floor in dBc/Hz as a function of number of
		resonators being read versus readout power per resonator for an amplifier
		noise temperature of 6\,K. The graph on the left corresponds to the
		logarithm of the contours \eqref{ruido_hemt_escalado} presented
		in~\cite{McHugh2012}, considering random phases, without specifying a
		generation algorithm. On the right the logarithm of the contours obtained
		using the algorithms presented in this paper are shown.  The dotted lines
		correspond to the use of Newman's phases, keeping the amplitudes constant.
		The solid lines represent the use of Newman phases in addition to the use of
		random amplitudes, that is, the most realistic case.}
	\label{fig_power_per_tones}
\end{figure*}
The same problem can be addressed from the point of view of the power applied to
the MKIDs. To get an idea of the improvement obtained when applying the Newman
phases, a comparison can be made between the results presented
in~\cite{McHugh2012} and the algorithms presented in this paper for the
estimation of the required phase noise floor in the ADC and the maximum power
applied to each MKID. \cite{McHugh2012} considers a signal consisting of $N$
tones generated with identical power $P_\text{rms}$ and random phases.
According to the hypotheses of that work, the crest factor can be approximated
by $CF\sim \frac{P} {NP_\text{rms}}$, where $P$ is the peak power of the $N$
tones signal. They mention the consequence of this hypothesis is that the power
gain available between MKIDs and ADCs is proportional to $\sqrt{N}$.  This can
be compared to the proportionality with $\sqrt{\log{N}}$ if the algorithms
presented in this paper are used. Since the contribution of the ADC to the
system noise (referred to the input of the MKID array) depends on the gain of
the system, 
the increase in the amount of signals in the FDM signal requires a lower noise
level from the ADC, which can be expressed in terms of bit noise, ie the
effective number of bits (ENOB).


\figurename~\ref{fig_power_per_tones} shows the comparison between
results of the maximum allowable power per tone in arrays of 100 to 10.000
resonators applied to a single DAC output. In calculating the contours of 
\figurename~\ref{fig_power_per_tones}, the noise scaling of the HEMT
amplifier is performed with a factor dependent on the number of channels, 
\begin{equation}\label{ruido_hemt_escalado}
	S = \frac{k_BT_\text{LNA}}{s_\text{max}^2 P_\text{rms}}
\end{equation}
and assuming a 3\,dB margin between the noise of the HEMT amplifier and the ADC.
$s_\text{max}$ sets the full scale width that the digitized multi-tone signal
can achieve. Furthermore, in calculating the contours of
\figurename~\ref{fig_power_per_tones}, a variation of up to 10\,dB in the power
$P_\text{rms}$ of each MKID is taken into account, considering both variations
constructive and of the material~\cite{McHugh2012}.

Analyzing the results presented in \figurename~\ref{fig_power_per_tones},
it can be concluded that for the algorithms used in this paper there is less
variation in the readout power per pixel through the entire range $N$.
In addition, the power per pixel is higher, which can result in a saving
of attenuators in the electronic readout chain.

\section{Conclusions}\label{sec:conc}
From the work presented, it is possible to understand the hardware requirements
of a system designed for the excitation and readout of large arrays of
superconducting resonators (detectors). It was shown the signal path through the electronic
readout, from the generation of the carriers to the responses of the detectors.

The main contribution be seen in the optimal method found to
generate the carriers. It was shown how the generation of phases using the
Newman algorithm leads to optimal results in terms of the crest factor of the
multi-frequency signal. Using the algorithms presented ensure less variation in
the reading power per pixel throughout the range N. The power per pixel is
greater, which translates into hardware savings in the readout system.

\appendix [Crest factor]\label{crest_factor}
The crest factor ($CF$) is defined as the ratio of the absolute value of the
peak in the data divided by the value of the root mean square (rms) of the same
data. 
This is a well-defined quantity for a deterministic process. However, in the
case of the waveform of a random variable, it cannot be defined with certainty.
Instead, it is commonly specified based on the percentage of times the maximum
value is greater (or less) than a specified value. For example, if a voltage $v$
has a crest factor that is greater than 2.576 only 1\% of the time. This
means, for example, that if we look at the waveform of $v$ on an oscilloscope
for 100 seconds, the total time the maximum value exceeds 2,576$v_\text{rms}$
is only 1 second out of 100.

Now, for a periodic signal $u$, of period $T$, the norms $L^\infty$ and
$L^2$ are defined by
\begin{equation}\label{norma_L_infinito}
  ||u||_\infty \triangleq \underset{t}{\text{sup}}|u(t)|
\end{equation}
and
\begin{equation}\label{norma_L_cuadrado}
  ||u||_2 \triangleq \left(\frac{1}{T}\int_0^T (u(t))^2 dt\right)^{1/2}.
\end{equation}
In engineering terms, $||u||_\infty$ is nothing other than the peak of the
signal $u$ and $||u||_2$ is its effective value. The \emph{crest factor} of a
non-zero signal is defined as the ratio of its peak value to rms value
\begin{equation}\label{def_factor_de_cresta}
  CF(u) \triangleq \frac{||u||_\infty}{||u||_2}.
\end{equation}
Normally the crest factor is expressed in decibels, that is,
\begin{equation}\label{def_factor_de_cresta_db}
  CF_{dB}(u) = 20 \log_{10}(CF(u)).
\end{equation}

An important characteristic is that the crest factor is always at least one.
This can be seen as follows: $u(t)^2\leq ||u||_\infty^2$ for all $t$, therefore
\begin{equation}\label{prueba_factor_de_cresta_igual_1}
  \frac{1}{T}\int_0^T (u(t))^2 dt \leq ||u||_\infty^2.
\end{equation}
Taking square roots and dividing gives $CF(u) \geq 1$.


%
%

%


\bibliographystyle{IEEEtran}
\bibliography{IEEEabrv, library, mybib, mybooks}

\end{document}